\documentclass[nofootinbib, reprint, preprintnumbers, amsmath, amssymb, amsfonts, aps, prx, superscriptaddress, floatfix]{revtex4-2}
\usepackage[utf8]{inputenc}
\usepackage{mathtools}
\usepackage{graphicx}
\usepackage{dcolumn}
\usepackage{bm}
\usepackage{physics}
\usepackage{color}
\usepackage{braket}
\usepackage[normalem]{ulem}
\usepackage{enumerate}
\usepackage{enumitem}
\usepackage[colorlinks=true,linkcolor=blue,citecolor=blue,urlcolor =blue]{hyperref}
\usepackage{mathrsfs}
\usepackage{bbm}

\usepackage{comment}
\usepackage{lipsum}

\begin{document}
\title{Hardware-Efficient Bosonic Quantum Computing with Photon-loss Detection Capability}

\author{Yuichiro Mori}
\email{mori-yuichiro.9302@aist.go.jp}
\affiliation{Global Research and Development Center for Business by Quantum-AI Technology (G-QuAT), National Institute of Advanced Industrial Science and Technology (AIST), 1-1-1, Umezono, Tsukuba, Ibaraki 305-8568, Japan}%

\author{Yuichiro Matsuzaki}
\email{ymatsuzaki872@g.chuo-u.ac.jp}
\affiliation{Department of Electrical, Electronic, and Communication Engineering, Faculty of Science and Engineering, Chuo university, 1-13-27, Kasuga, Bunkyo-ku, Tokyo 112-8551, Japan}%

\author{Suguru Endo}
\email{suguru.endou@ntt.com}
\affiliation{NTT Computer and Data Science Laboratories,
NTT Corporation, Musashino, 180-8585, Tokyo, Japan}

\author{Shiro Kawabata}
\email{s-kawabata@aist.go.jp}
\affiliation{Global Research and Development Center for Business by Quantum-AI Technology (G-QuAT), National Institute of Advanced Industrial Science and Technology (AIST), 1-1-1, Umezono, Tsukuba, Ibaraki 305-8568, Japan}%

\affiliation{NEC-AIST Quantum Technology Cooperative Research Laboratory,
National Institute of Advanced Industrial Science and Technology (AIST), 1-1-1, Umezono, Tsukuba, Ibaraki 305-8568, Japan}

\date{\today}

%%%%%%%%

\begin{abstract}
Bosonic quantum systems offer the hardware-efficient construction of error detection/error correction codes by using the infinitely large Hilbert space. However, due to the encoding, arbitrary gate rotations usually require magic state teleportation or complicated optimized pulse sequences involving an ancilla qubit. Here, we propose a simple and hardware-efficient bosonic $02$ error detection code that allows for the implementation of arbitrary $X$ and $Z$ rotations and a controlled phase gate by using a Kerr nonlinear resonator. Our code can detect a single photon loss, and we observe significant error suppression by simulating the frequently used hardware-efficient ansatz quantum circuit in near-term quantum computing. 
\end{abstract}

\maketitle

%%%%%%%%%%
\section{Introduction}
\label{sec:intro}
%%%%%%%%%%
Quantum error correction (QEC) and quantum error detection (QED) play a crucial role in the field of quantum information theory~\cite{Peres1985PRA, nielsen2010quantum, Terhal2015RMP, SuterRMP2016, Roffe_2019CP} to protect the quantum information from the detrimental effect of environmental noise. While the code words of conventional quantum error correction/detection codes are built from many physical qubits, bosonic quantum error correction codes exploit the infinite-dimensional Hilbert space of bosonic codes to extract the useful energy levels for QEC or QED~\cite{VAlbert_2018PRA, Grimsmo_2020PRX, Hillmann_2022PRXQ, Tsunoda_2023PRXQ, Mizuno_2023, Schlegel_2022PRA, Fukui_2023PRA, Royer_2022PRXQ}. Because we do not use multiple physical systems, this error correction strategy is much more hardware-efficient and circumvents engineering problems to connect many qubits.

Because of the hardware efficiency of bosonic QEC/QED codes, they may offer robust error suppression in near-term quantum hardware. However, the implementation of bosonic codes incurs significant hardware overheads. For example, while many near-term algorithms, e.g., variational quantum algorithms, assume the capability of applying arbitrary rotation angles in variational quantum circuits~\cite{Bharti_22_RMP,cerezo2021variational}, we need to teleport the magic states to apply gate operations with arbitrary rotation angles fault-tolerantly in general encoded scenarios~\cite{grimsmo2020quantum,grimsmo2021quantum}. Meanwhile, another strategy is to obtain arbitrary operations with pulse operations consisting of a mixture of displacement operations and controlled rotation operations~\cite{heeres2017implementing}; however, this method involves complex pulse optimization and an ancilla qubit.

In this paper, we propose a quite hardware-efficient code tailored for bosonic NISQ computers, enabling the implementation of gate operations through straightforward pulse operations without ancillary quantum systems. This code allows for detecting single-photon losses, a critical noise in the superconducting qubits. We use one of the $0N$ code~\cite{Sabapathy_2018PRA, Elder_2020PRX, Grimsmo_2020PRX}, namely,
\begin{align}
    |0_{L}\rangle=|0\rangle,\quad |1_{L}\rangle = |2\rangle,\label{eq:code}
\end{align}
where the subscript $L$ denotes the logical state. We call this code $02$ code in this paper. 
This code allows us to perform the detection of the single photon loss. Let us consider the measurement of a parity operator $(-1)^{\hat{a}^{\dagger}\hat{a}}$, where $\hat{a}^{\dagger}$ ($\hat{a}$) represents a creation (destruction) operator. If the measurement yields a result of $+1$, it confirms that the quantum state is within the logical-qubit subspace. Conversely, if the measurement yields the other result, it indicates leakage from the logical subspace, prompting us to discard the state. We also show that the two-photon parametric driving allows for the execution of single-qubit operations along the $x$-axis, while rotations along the $z$-axis can be achieved through detuning. Additionally, we simulate a controlled-phase gate with adiabatically controlling the coupling strength. Combining these gates, we prepare a quantum ansatz circuit and evaluate the fidelity of the output state. Lastly, we verify the effectiveness of the single photon loss detection using the ansatz circuit.

The remainder of this paper is as follows. In Sec.~\ref{sec:Gate_operations}, we explain how to implement a universal gate set on logical qubits encoded by the 02 code without any ancillary systems. In Sec.~\ref{sec:detect}, we explain how to detect the single-photon loss. In Sec.~\ref{sec:simulation}, we present the numerical results of our proposed method. Finally, we offer some concluding remarks in Sec.~\ref{sec:conc_disc}. Throughout the paper, $\hbar =1$.

\section{Gate implementation and readout}
\label{sec:Gate_operations}
%%%%%%%%%%
We consider Kerr non-linear resonators (KNRs)~\cite{Milburn1991PRA,Wielinga1993PRA,Cochrane1999PRA}, which can be realized using superconducting circuits~\cite{Goto_2016srep, Puri2017_npjq,Meaney2014EPJQ,WangPRX2019,Grimm2020_nature,Yamaji2022PRA}. The Hamiltonian is as follows:
\begin{align}
    \hat{H} = \sum_{i}&\left(K_{i} \hat{a}_{i}^{\dagger 2}\hat{a}_{i}^{2} + \Delta_{i} \hat{a}_{i}^{\dagger} \hat{a}_{i} + p_{i} (\hat{a}_{i}^{2} + \hat{a}_{i}^{\dagger 2})\right)\nonumber \\
    & +\sum_{i>j}g_{ij}(\hat{a}_{i}^{\dagger}\hat{a}_{j} + \hat{a}_{i}\hat{a}^{\dagger}_{j}),\label{eq:Hamiltonian}
\end{align}
where $K_{i}$ is the Kerr nonlinearity, $\Delta$ is the detuning between the driving and cavity fields,  $p_{i}$ is the amplitude of the parametric derive, and $g_{ij}$ is the coupling strength of the interaction between two KNRs. On the other hand, we do not bifurcate the system, and we just use the parametric drive to implement the $X$ rotation in our method.

We assume that the coupling strength can be dynamically changed. Such a control of the coupling strength can be realized by using superconducting circuits~\cite{Pfaff_2017NatPhy,yan2018tunable,stehlik2021tunable}.
Notably, we do not use the coherent drive, described by $\hat{a}_i +\hat{a}_i^{\dagger}$, which is usually used in manipulating the state of superconducting qubits. To confine the state in the logical subspace, we use the parametric drive instead of the coherent drive.

%%%%%%
\subsection{Single-qubit gates}
\label{ssec:single-qubit}
%%%%%%
First, let us consider the single-qubit rotation. 
We assume that, during the implementation of the single qubit rotations, we turn off the interaction.
To perform the $Z$ rotation, we set $p_{i} = 0$, and the Hamiltonian is described as
\begin{align}
    \hat{H}=K \hat{a}^{\dagger 2}\hat{a}^{2} + \Delta \hat{a}^{\dagger}\hat{a}.\label{Ham_Rams}
\end{align}
Here, by choosing $\Delta \neq -K$, we can set a finite detuning between $\ket{0}$ and $\ket{2}$.
Then, the transition matrix is 
\begin{align}
    \begin{pmatrix}
        \langle 0_{L}|\hat{H}|0_{L}\rangle & \langle 1_{L}|\hat{H}|0_{L}\rangle \\
        \langle 0_{L}|\hat{H}|1_{L}\rangle & \langle 1_{L}|\hat{H}|1_{L}\rangle
    \end{pmatrix}
    =\begin{pmatrix}
        0 & 0 \\
        0 & 2K+2\Delta
    \end{pmatrix}.
\end{align}

When we perform the $Z$ rotation by an angle of $\theta$, we let the state evolve by the Hamiltonian for a time of $\tau_{Z}$ where $\tau_{Z}$ satisfies
\begin{align}
    2(K + \Delta)\tau_{Z} = \theta.\label{Z_gate_duration}
\end{align}

To perform the $X$ rotation, we use the following Hamiltonian
\begin{align}
    \hat{H} = K \hat{a}^{\dagger 2}\hat{a}^{2} -K \hat{a}^{\dagger} \hat{a} + p (\hat{a}^{2}+\hat{a}^{\dagger 2}).\label{Ham_Rabi}
\end{align}
Then, the transition matrix is 
\begin{align}
    \begin{pmatrix}
        \langle 0_{L}|\hat{H}|0_{L}\rangle & \langle 1_{L}|\hat{H}|0_{L}\rangle \\
        \langle 0_{L}|\hat{H}|1_{L}\rangle & \langle 1_{L}|\hat{H}|1_{L}\rangle
    \end{pmatrix}
    =\begin{pmatrix}
        0 & \sqrt{2}p \\
        \sqrt{2}p & 0
    \end{pmatrix}.
\end{align}
By setting $K \ll p$, the dynamics induced by the Hamiltonian is confined in a subspace spanned by $|0\rangle $ and $|2\rangle $ effectively. 

The rotation angle $\theta$ is determined by the duration $\tau_{X}$ as
\begin{align}
\sqrt{2}p\tau_{X} = \frac{\theta}{2}.\label{X_gate_duration}
\end{align}

%%%%%%
\subsection{Two-qubit gate (Controlled-phase gate)}
\label{ssec:cphase-qubit}
%%%%%%
Let us explain how to perform the controlled-phase gate between two KNRs.
The Hamiltonian is described as follows:
\begin{align}
    \hat{H} = &\sum_{i=1,2}\left(K_{i} \hat{a}_{i}^{\dagger 2}\hat{a}_{i}^{2} + \Delta_{i} \hat{a}_{i}^{\dagger} \hat{a}_{i} \right)\nonumber \\
    &\quad +  g (\hat{a}_{1}^{\dagger}\hat{a}_{2}+\hat{a}_{2}^{\dagger}\hat{a}_{1}).\label{CZ:Ham}
\end{align}
As we describe in Appendix~\ref{app:enederiv}, the effective Hamiltonian describing the time evolution induced by the beam-splitter interaction contains terms of $\hat{n}_{1}\hat{n}_{2}$, which appears in the second-order perturbation. Such an interaction has been used for qutrit entangling gates with superconducting transmon qubits~\cite{goss2022high}. On the other hand, we utilize it to implement the controlled phase gate for the 02 code. 

To obtain this effective interaction, we need a condition $E_{02}\neq E_{20}$, or equivalently, $K_{1}\neq K_{2}$ or $\Delta_{1}\neq\Delta_{2}$. We define an energy difference as
\begin{align}
    \delta E = |E_{22}-(E_{02} + E_{20})|.\label{eq:endif}
\end{align}
As seen in Eq.~\eqref{pert_hosei_ener}, the energy difference depends on the coupling constant $g$ as follows:
\begin{align}
     &\delta E(g)\nonumber\\
     &= -\frac{6g^{2}}{4K_{1}-2K_{2}+\Delta_{1}-\Delta_{2}}-\frac{6g^{2}}{4K_{2}-2K_{1}-\Delta_{1}+\Delta_{2}}\nonumber\\
     &\qquad -\frac{2g^{2}}{2K_{2}-\Delta_{1}+\Delta_{2}}- \frac{2g^{2}}{2K_{1}+\Delta_{1}-\Delta_{2}},\label{gdepdelE}
\end{align}
where $K_1$ and $K_2$ represent the Kerr coefficients of the first and second KNRs, while $\Delta_1$ and $\Delta_2$ represent the detuning of the first and second KNRs, respectively. 

We assume that the initial state is given by
\begin{align}
\ket{\psi} = c_{00}\ket{00}+c_{02}\ket{02}+c_{20}\ket{20}+c_{22}\ket{22}.
\end{align}
Then, the state after time $t$ under the Hamiltonian~\eqref{CZ:Ham} is
\begin{align}
\ket{\psi(t)} = &c_{00}\ket{00}+c_{02}e^{-iE_{02}t}\ket{02}\nonumber \\
&\ +c_{20}e^{-iE_{20}t}\ket{20}+c_{22}e^{-iE_{22}t}\ket{22}.\label{eq:state_CZga}
\end{align}

To perform the controlled-\textit{Z} rotation, we set the rotation angle with the condition,
\begin{align}
\delta E t &= \theta. \label{eq:cond_CZ}
\end{align} 
We can control the phase $e^{-iE_{02}t}$ and $e^{-iE_{20}t}$ by performing the single-qubit $Z$ rotation. Then, we obtain,
\begin{align}
    \ket{\psi(t)} = c_{00}\ket{00}+c_{02}\ket{02}+c_{20}\ket{20}+e^{-i\theta}c_{22}\ket{22}.\label{state_cphase}
\end{align}

When adjusting the coupling $g$, it is typically handled smoothly over time~\cite{Grimm2020_nature}. We consider the case that we can change the coupling strength $g$ during the gate operation. If the change is slow enough to satisfy an adiabatic condition, the adiabatic theorem~\cite{Born_Fock_1928, KatoT_1950_JPSJ, messiah1999quantum} shows that we obtain the state in Eq.~\eqref{state_cphase}, when the following conditions are satisfied:
\begin{align}
    \int_{t_\mathrm{i}}^{t_{f}}\delta E(t) dt &= \theta,\nonumber\\
    g(t_{i}) = g(t_{f}) = 0,&\quad t_{f}-t_{i}\gg \frac{g(t)}{E_{02}^{2}(t)}, \frac{g(t)}{E_{20}^{2}(t)},\label{condition_CPhase}
\end{align}
Here, $E_{02}(t)$ and $E_{20}(t)$ are instantaneous eigenvalues, $t_{i}$ is the initial time of the operation, $t_{f}$ is the final time of the operation. The last condition assures the adiabaticity.

\subsection{Readout}
\label{subsec:readout}
Due to the small energy of microwave photons, we rely on amplification of weak microwave signals to perform a measurement~\cite{wallraff2005approaching, Hann2018PRA,Blais_2021RMP}. 
Existing readout technology uses the dispersive coupling,
\begin{align}
\hat{H}=g \hat{n}_{1}\hat{n}_{2},    
\end{align}
where $g$ is a coupling constant, $\hat{n}_{1}$ ($\hat{n}_{2}$) is the number operator of the KNR (ancillary system to be used for the readout). Therefore, it is expected that the readout can be performed on the qubits encoded by the 02 code with the same architecture, and it has already been realized experimentally~\cite{Elder_2020PRX,Kohler2018PRA}. So, in our simulation, we assume that the readout can be ideally performed.

%%%%%
\subsection{Simulation of the gates}
\label{Sim:gate}
%%%%%
We perform numerical simulations of the three individual gates, $X$ rotation, $Z$ rotation, and controlled-Z gate. Also, we evaluate the fidelity between the target states and states obtained by solving the Schr\"odinger~\cite{schroedinger1926_pr} and the Gorini–Kossakowski–Sudarshan–Lindblad (GKSL) master equations~\cite{GKS_1976,Lindblad_1976}. The GKSL master equation for the density matrix $\rho$ is described as
\begin{align}
    \dot{\rho} = -i[\hat{H},\rho] + \sum_{i}\left(\hat{L}_{i}\rho\hat{L}^{\dagger}_{i}-\frac{1}{2}\{\hat{L}^{\dagger}_{i}\hat{L}_{i},\rho\}\right),\label{GKSL}
\end{align}
where $\hat{H}$ is the Hamiltonian, $\hat{L}_{i}=\sqrt{\gamma}\hat{a}_{i}$ are the Lindblad operators to describe a single-photon loss. We chose $\gamma = 0.002\mathrm{MHz}$, which refers to the experimentally obtained value with a superconducting circuit~\cite{wang2022towards}.

To illustrate the potential capability of improving the fidelity, we start from a simplified scenario where we can apply an ideal parity measurement onto the logical space to the state evolved by the GKSL equation to perform the error detection. We will consider the imperfection of such parity measurements in the next section.

\begin{figure}
    \includegraphics[width = 8cm]{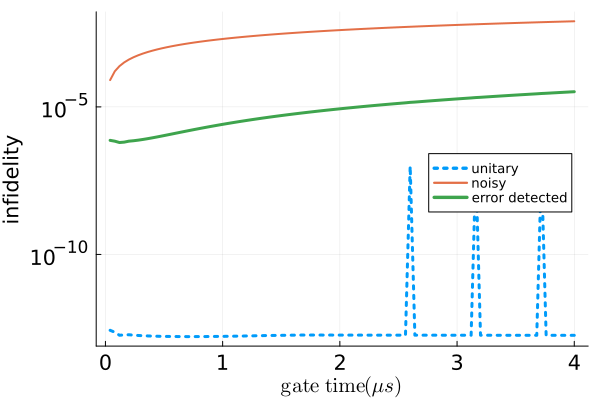}
    \caption{Relationship between the infidelity of states after passing through a $Z$ gate and gate time is depicted for three scenarios: noise-free (unitary), noisy without error detection (noisy), and error detected (error detected) cases. While theoretically zero in the noise-free case, numerical errors result in non-zero values. The figure displays several points with significant numerical errors as observed.}\label{Z_rot}
\end{figure}

\begin{figure}
    \includegraphics[width = 8cm]{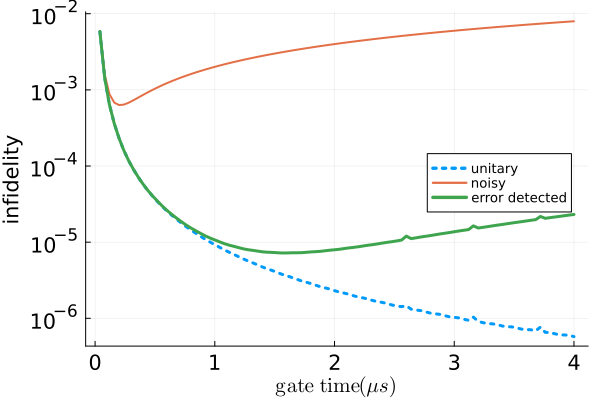}
    \caption{Infidelity of states after passing through an X gate versus gate time for the three cases. In the case of the X gate, even under noise-free case, there are slight transitions to other excited states such as $\ket{4}$, resulting in non-zero infidelity. Thus, a trade-off relationship with errors due to single-photon loss is observed.}\label{X_rot}
\end{figure}

We explain the simulation of the $Z$ and $X$ rotations where we use the Hamiltonian in Eq.~\eqref{Ham_Rams} (Eq.~\eqref{Ham_Rabi}). In the $Z$ ($X$) rotation, we set the initial state to the state $|0\rangle$ ($|+\rangle = (|1\rangle_{L}+|0\rangle_{L})/\sqrt{2}$) and calculate the fidelity with the state $|1\rangle$ ($|-\rangle = (|0\rangle_{L}-|1\rangle_{L})/\sqrt{2}$). The parameters we set are $K=250\ \mathrm{MHz}$, $\Delta=-248.43 \mathrm{MHz}$ (in $Z$ rotation), and $p = 1.11 \mathrm{MHz}$ (in $X$ rotation). These parameters are chosen with referring to Ref.~\cite{Bourassa2012PRA,ChonoPRR2022}. The results are depicted in Fig.~\ref{Z_rot} and \ref{X_rot}. While the fidelity is decreased by decoherence caused by single-photon loss in both cases, the fidelity is improved by detecting single-photon loss.
Remarkably, the optimal time to achieve the maximum fidelity with quantum error detection is longer than that without quantum error detection in the $X$ rotation case, as we explain the reason of this in Appendix~\ref{appendix_B}.


Next, we explain the simulation of the controlled-phase gate where we use the Hamiltonian in Eq.~\eqref{CZ:Ham}. In our approach, we need to compensate a redundant phase by using the $Z$ rotation of a single qubit for the two modes. Therefore, after determining the required correction angles, we apply corrections for each mode using the aforementioned single-qubit $Z$-rotations. For the control of the interaction strength $g$, we assume the following function:
\begin{align}
    g(t) = g_{0}\left(\frac{2}{\pi}\right)^{\frac{1}{4}}\mathrm{exp}\left[-\frac{(t-t_{0})^{2}}{\tau^{2}}\right],\label{couple_time_func}
\end{align}
where a parameter $\tau$ satisfies a condition 
\begin{align}
    \int_{t_{i}}^{t_{f}}\delta E(g(t)) dt = \theta.\label{g_t}
\end{align}
Here, $\delta E(g)$ is given by Eq.~\eqref{gdepdelE}. Subsequently, by adjusting the detuning, we perform single-qubit rotations for each KNR.

Suppose we implement the gate from $t=0$ to $t=2t_0$. The left hand side of Eq.~\eqref{g_t} is calculated as
\begin{align}
    \int_{t_{i}}^{t_{f}}\delta E(g(t))dt &=\delta E(g_{0})\tau\times \mathrm{erf}\left[\sqrt{2}\frac{t_{0}}{\tau}\right].
\end{align} 
In our simulation, we set $t_{0} = 3\tau$ and $\mathrm{erf}\left[3\sqrt{2}\right]\simeq 1$ is satisfied.

We chose $K_{1}=250 \mathrm{MHz}$, $K_{2}=200 \mathrm{MHz}$, $\Delta_{1}=-250 \mathrm{MHz}$,  $\Delta_{2}=-170 \mathrm{MHz}$, and $g_{0}=15 \mathrm{MHz}$. The initial state is
\begin{align}
    |\psi\rangle = \frac{1}{2}(|00\rangle+|20\rangle+|02\rangle+|22\rangle)
\end{align}
in this simulation and the fidelity of the state after the CZ operation ($\theta = \pi$) is shown by Table~\ref{tab:fid_CZ}. The impact of decoherence can be significantly mitigated (fidelity $= 0.98877 \to 0.99988$) through the detection, as well as the single-qubit rotations.
\begin{table}[h]
    \centering
    \caption{Fidelity of CZ gate. Even in the absence of noise, the fidelity does not reach $1$ due to non-adiabatic transitions. }\label{tab:fid_CZ}
    \begin{tabular}{|c|c|c|}
    \hline
        Unitary &  W/o error detection & With error detection\\
        \hline
       $0.999960$ & $0.987634$ & $0.999912$ \\
       \hline
    \end{tabular}
\end{table}
When the single-photon loss occur twice during each gate operation, the impact of single-photon loss will remain if our quantum error detection is performed. To address this issue, either continuously monitor single-photon losses during gate operations or, alternatively, employ other methods such as quantum error mitigation.

%%%%%%%%%
\section{Detection of a single-photon loss}
\label{sec:detect}
%%%%%%%%%
This section explains a method to perform the number parity measurement for KNRs. 
%%%%%
\subsection{Single-photon loss detection}
\label{erdtsc}
%%%%%
Here, we provide a method to detect single-photon loss. First, we use an ancillary qubit such as a superconducting transmon qubit, and prepare the state,
\begin{align}
    |+\rangle = \frac{1}{\sqrt{2}}(\ket{\downarrow} + \ket{\uparrow}).
\end{align}
We use an interaction between the KNR and this ancillary qubit is written as,
\begin{align}
    \hat{H}_{det} = \chi \hat{a}^{\dagger}\hat{a} \ket{\uparrow}\bra{ \uparrow}.\label{det_Ham}
\end{align}
to perform the parity measurement~\cite{besse2020parity,von2022parity}. As we describe in Appendix~\ref{app:enederiv}, we can obtain this Hamiltonian from the following model,
\begin{align}
\hat{H} = K\hat{a}^{\dagger 2}\hat{a}^{2}+\Delta\hat{a}^{\dagger}\hat{a}+\frac{E}{2}\sigma_{Z}+g'(\hat{a}\sigma_{+}+\hat{a}^{\dagger}\sigma_{-}),\label{Hamil_coup}
\end{align}
where we use the dispersive approximation~\cite{walter2017rapid,siddiqi2006dispersive,iyama2024observation}.
It is worth mentioning that, while a linear harmonic oscillator was typically coupled with a qubit~\cite{walter2017rapid,siddiqi2006dispersive}, we consider a case that the KNR is coupled with a qubit~\cite{iyama2024observation}. We discuss the condition for the dispersive approximation for such a case in Appendix~\ref{app:enederiv}. 

From the the simplified Hamiltonian represented by Eq.~\eqref{det_Ham}, we obtain the following unitary operator,
\begin{align}
    \hat{U}(t) &= e^{-i\hat{H}_{det}t},\nonumber\\
    &=e^{-it\chi \hat{a}^{\dagger}\hat{a}}\ket{\uparrow}\bra{\uparrow}+\hat{I}\ket{\downarrow}\bra{\downarrow}.
\end{align}
We choose $t$ such that $\chi t = \pi$, the unitary operator becomes,
\begin{align}
\hat{U}(t) &= (-1)^{\hat{a}^{\dagger}\hat{a}}\ket{\uparrow}\bra{ \uparrow}+\hat{I}\ket{\downarrow}\bra{\downarrow},\nonumber \\
&=\frac{\hat{I}+(-1)^{\hat{a}^{\dagger}\hat{a}}}{2}\hat{I}-\frac{\hat{I}-(-1)^{\hat{a}^{\dagger}\hat{a}}}{2}\hat{Z},\label{coup_unitary}
\end{align}
where $\hat{Z}=\ket{\uparrow}\bra{\uparrow}-\ket{\downarrow}\bra{\downarrow}=\ket{+}\bra{-}+\ket{-}\bra{+}$.
After the dynamics, we perform the projective $X$ measurement on the ancillary qubit. When the measurement result is $+1$, we successfully project the fock states into even-number subspace. So, we obtain 
\begin{align}
    \rho' = \hat{P}\rho\hat{P},
\end{align}
where $\hat{P}$ is defined by
\begin{align}
    \hat{P}=\frac{\hat{I}+(-1)^{\hat{a}^{\dagger}\hat{a}}}{2}.\label{parity_project}
\end{align}
When we use multiple KNRs, we can perform this parity measurement on each KNR, and we obtain 
\begin{align}
\rho_{ed}&=\hat{\Pi}\rho\hat{\Pi}.\label{def:rho_ed}
\end{align}
Here, $\hat{\Pi}$ is the projector onto the logical subspace.

%%%%%%
\subsection{Simulation of the detection}
\label{sim_detect}
%%%%%%
To validate our scheme of single-photon loss detection, we perform simulations and evaluate the process fidelity of the parity measurement. We use the Hamiltonian~\eqref{Hamil_coup}, and the GKSL equation defined by Eq.~\eqref{GKSL} with $\hat{L}_{1}=\sqrt{\gamma}\hat{a}$ and $\hat{L}_{2}=\sqrt{\gamma}\sigma_{-}$.  We utilized $K=250\mathrm{MHz}$, $\Delta = 250\mathrm{MHz}$, $E=\Delta + K = 500\mathrm{MHz}$, and $\gamma = 0.002\mathrm{MHz}$ in this simulation. To compare the case without the single-photon loss, we also perform a simulation with a condition $\gamma = 0$.
As we considered the controlled-$Z$ gate in Sec.~\ref{sec:simulation}, we assume the coupling $g'$ in Eq.~\eqref{Hamil_coup} is adjustable and the time dependence is given by $g'(t) = g'_{0}\left(\frac{2}{\pi}\right)^{\frac{1}{4}}\mathrm{exp}\left[-\frac{(t-t_{0})^{2}}{\tau'^{2}}\right]$,
while $\tau'$ and $g'_{0}$ satisfy the condition,
\begin{align}
    2\frac{g_{0}^{'2}}{K}\tau = \pi.
\end{align}
In the time evolution induced by this time dependent coupling $g'$, non-adiabatic transitions may be caused. 

To consider the validity of the process, we evaluate the process fidelity, which is defined as follows~\cite{Gilchrist_2005, NIELSEN2002249, Greenaway_2021PRR, BOWDREY2002258}:
\begin{align}
    F(\Lambda, \Gamma) = \frac{1}{d^{2}} \sum_{i=1}^{d^{2}}\mathrm{Tr}[\Lambda(\sigma_{i}^{\dagger})\Gamma(\sigma_{i})],
\end{align}
where $\Lambda$ is a desired channel that maps input states, $\Gamma$ is a realized channel that does not perfectly coincide with $\Lambda$ and $\sigma_{i}$ compose a complete set of mutually orthonormal operators on a $d$-dimensional Hilbert space. 
In our case, the desired channel $\Lambda$ is given by the unitary operation Eq.~\eqref{coup_unitary} and the realized channel $\Gamma$ is given by the unitary operator induced by the Hamiltonian~\eqref{Hamil_coup} with adjusting the relative phase on each mode.
We consider that the dynamics should be confined within the logical subspace whose dimension is $d=2$ so that we choose a complete set of mutually orthonormal operators on the logical subspace, \textit{i.e.,}
\begin{align}
    \sigma_{1}&=\frac{1}{\sqrt{2}}(\ket{0}\bra{0}+\ket{2}\bra{2})\otimes \ket{+}\bra{+}, \\
    \sigma_{2}&=\frac{1}{\sqrt{2}}(\ket{2}\bra{0}+\ket{0}\bra{2})\otimes \ket{+}\bra{+},\\
    \sigma_{3}&=\frac{1}{\sqrt{2}}(-i\ket{2}\bra{0}+i\ket{0}\bra{2})\otimes \ket{+}\bra{+},\\
    \sigma_{4}&=\frac{1}{\sqrt{2}}(\ket{0}\bra{0}-\ket{2}\bra{2})\otimes \ket{+}\bra{+}.
\end{align}
The process fidelity is $0.998$ when there is no decoherence ($\gamma=0$) and $0.991$ ($\gamma=0.002\ \mathrm{MHz}$).

%%%%%%%%%
\section{Simulation with an ansatz circuit}
\label{sec:simulation}
\begin{figure}
    \centering
    \includegraphics[width = 8.5cm]{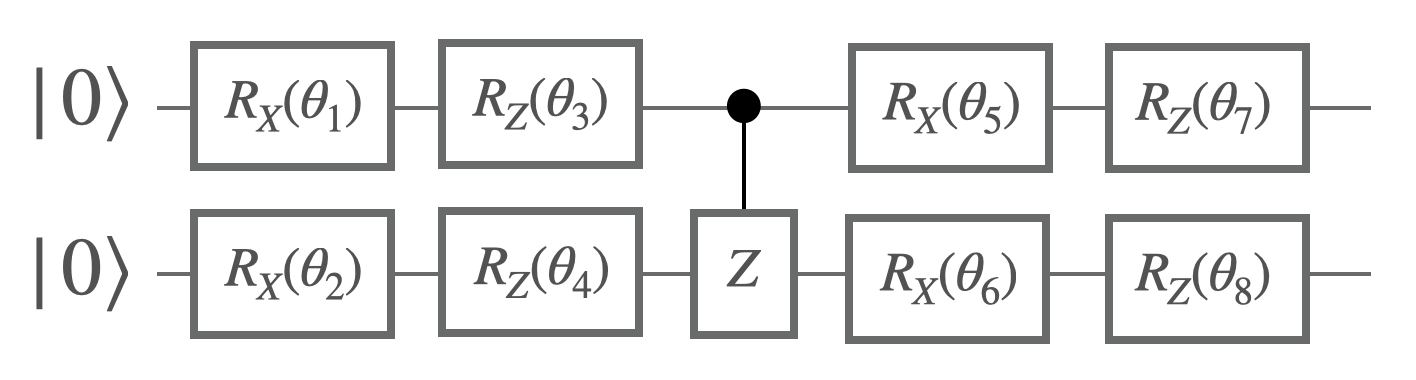}
    \caption{Ansatz quantum circuit for the simulation. It is composed of single-qubit rotation gates and the controlled-$Z$ gates with two qubits.}
    \label{fig:HEA}
\end{figure}

%%%%%%%%%
In this section, we evaluate the performance of our method for NISQ computing. To implement variational algorithms, it is necessary to prepare a parametrized quantum circuit called an \textit{ansatz} circuit. We consider a circuit with $8$ parameters with our proposed gates, as shown in Fig.~\ref{fig:HEA}. We perform numerical simulations with this circuit while we perform quantum error detection by using our method.

Let us explain the parameters to perform the gate operations in this section. The Hamiltonian in this simulation is
\begin{align}
    \hat{H}&=\sum_{i=1,2}\left(K_{i}\hat{a}^{\dagger 2}_{i}\hat{a}^{2}_{i}+\Delta_{i}\hat{a}^{\dagger}_{i}\hat{a}_{i}+p_{i}(\hat{a}^{\dagger 2}_{i}+\hat{a}^{2}_{i})\right)\nonumber\\
    &\qquad+g_{12}(t)(\hat{a}^{\dagger}_{1}\hat{a}_{2}+\hat{a}_{1}\hat{a}^{\dagger}_{2})\nonumber\\
    &\qquad +\sum_{i=1,2}\left(\frac{E}{2}\sigma^{(i)}_{Z}+g'_{i}(t)(\hat{a}_{i}\sigma^{(i)}_{+}+\hat{a}^{\dagger}_{i}\sigma^{(i)}_{-})\right).
\end{align}
Here, $g_{12}(t)$, $g'_{1}(t)$, and $g'_{2}(t)$ have the same time-dependence form given by Eq.~\eqref{couple_time_func}, while the strength is different.
We have
$g_{12}(t) = g_{120}\left(\frac{2}{\pi}\right)^{\frac{1}{4}}\mathrm{exp}\left[-\frac{(t-t_{0})^{2}}{\tau^{2}}\right]$, $g'_{1}(t) = g'_{10}\left(\frac{2}{\pi}\right)^{\frac{1}{4}}\mathrm{exp}\left[-\frac{(t-t_{0})^{2}}{\tau^{2}}\right]$, and 
$g'_{2}(t) = g'_{20}\left(\frac{2}{\pi}\right)^{\frac{1}{4}}\mathrm{exp}\left[-\frac{(t-t_{0})^{2}}{\tau^{2}}\right]$.
The parameters $K_{i}, \Delta_{i}$, $p_{i}$ $g_{012}$, $E$, and $g_{0i}$ for the simulations are shown in Table~\ref{tab:params_ansim}. Also, the Lindblad operators used in simulations in this section are $\sqrt{\gamma}\hat{a}_{1},\sqrt{\gamma}\hat{a}_{2}, \sqrt{\gamma}\hat{\sigma}_{-}^{(1)},$ and $\sqrt{\gamma}\hat{\sigma}_{-}^{(2)}$, where $\gamma = 0.002\ \mathrm{MHz}$  except for the unitary case ($\gamma = 0$).

\begin{table*}
    \centering
    \caption{Parameters we used in the simulation with the ansatz shown in Fig.~\ref{fig:HEA}. Single qubit rotations $R_{X}, R_{Z}$ can be performed at the same time on multiple modes. We give some parameter sets whose  corresponding operation is the product of the single qubit rotation on each mode. The unit of these numbers is $\mathrm{MHz}$.}\label{tab:params_ansim}
    \begin{tabular}{|c|c|c|c|c|c|c|c|c|c|c|}
    \hline
       Operation& $K_{1}$ &  $K_{2}$ & $\Delta_{1}$ & $\Delta_{2}$ & $p_{1}$ & $p_{2}$ & $g_{012}$ & $E$ & $g'_{01}$ & $g'_{02}$\\
        \hline
        $R_{X}(\theta_{1})\otimes R_{X}(\theta_{2})$ &$250$ & $250$ & $-250$ & $-250$ & $\theta_{1}$ & $\theta_{2}$ & $0$ & $0$ & $0$  &$0$\\
       \hline
       $R_{Z}(\theta_{3})\otimes R_{Z}(\theta_{4})$ &$250$ & $250$ & $-250-\theta_{3}$ & $-250 -\theta_{4}$ & $0$ & $0$ & $0$ & $0$ & $0$  &$0$\\
       \hline
       CZ &$250$ & $250$ & $-250$ & $-170$ & $0$ & $0$ & $15$ & $0$ & $0$  &$0$\\\hline
       Error detection & $250$ & $250$ & $250$ & $250$ & $0$ & $0$ & $0$ & $500$ & $25$ & $25$\\ \hline
    \end{tabular}
\end{table*}

\begin{figure}
    \centering
    \includegraphics[width = 8.5cm]{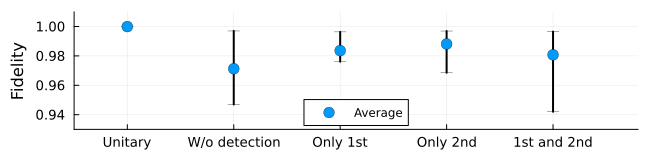}
    \caption{Fidelity between the ideal output of the ansatz and the actual outputs. }
    \label{fig:ansatz_fidelity}
\end{figure}

We evaluate the fidelity between the state obtained through the desired ansatz and the state obtained through our gates with $80$ randomly generated parameter sets, as shown in Fig.~\ref{fig:ansatz_fidelity}. In this calculation, we considered five different cases: The first case is that there is no single-photon loss (``Unitary'' in Fig.~\ref{fig:ansatz_fidelity}). 
Due to the approximation such as adiabaticity, the fidelity does not become unity, even if there is no decoherence. The second case is that single-photon loss occurs and we do not perform any detection of single-photon loss (``w/o detection''). In the third case and after, we perform the detection of single-photon loss. In the third case, we perform the detection just after the first controlled-$Z$ gate layer (``Only 1st''). In the fourth case, we perform the detection just after the last $Z$ rotation layer (``Only 2nd''). In the fifth case, we perform the detection at both layers (``1st and 2nd''). In the unitary case, the fidelity is almost $1$. Even the minimum value among $80$ trials is $0.99974$ in our simulation, which shows the validity of our approximation.

\begin{figure}
    \centering
    \includegraphics[width = 8cm]{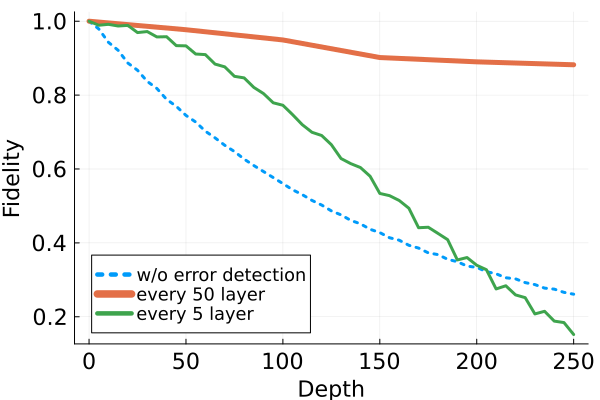}
    \caption{Fidelity against the depth of the quantum circuit repeating the ansatz shown in Fig.~\ref{fig:HEA} without resetting the state to the initial $|000\rangle$.}
    \label{fig:fidelity}
\end{figure}

In the cases ``Only 1st'' and ``Only 2nd'', we see that the fidelity becomes larger than the case of ``w/o detection''. This demonstrates that our method is useful to suppress the decoherence for quantum circuits. However, since our parity measurements are noisy to induce unwanted errors, many quantum error detections decrease the fidelity.
Let us consider a deeper quantum circuit to evaluate the impact of our imperfect parity measurements. We extend the number of layers by iteratively applying the circuit described in Fig.~\ref{fig:HEA}. We explore three cases: no error detection, error detection at every 5 layers, and error detection at every 50 layers, to investigate how the fidelity changes. The simulation results are shown in Fig.~\ref{fig:fidelity}. As we increase the depth, the fidelity decreases due to the accumulation of errors. The fidelity of the ``every 50 layers'' case is better than that of the ``w/o error detection'' case. However, the fidelity of the ``every 5 layers'' case is worse than that of the ``every 50 layers'' case. When the depth exceeds $200$, the ``every 5 layers'' case becomes worse than the ``w/o error detection''. In practical schemes, error detection itself can be a source of errors. Therefore, frequent error detection may lead to errors that are more pronounced than the single-photon losses.

%%%%%%%%%
\section{Conclusion and Discussions}
\label{sec:conc_disc}
%%%%%%%%%
We propose a method to implement a universal gate set with the 02 code without the need for ancillary qubits, and show that we can suppress the decoherence through the detection of single-photon losses. We use a Kerr non-linear resonator (KNRs) as a logical qubit, and we can perform the $X$ rotation and $Z$ rotation for the logical qubits by using a parametric drive and control of detuning, respectively. On the other hand, we can implement a controlled-phase gate by modulating the coupling strength of a beam-splitter type interaction. 
We further show that our proposed gates exhibit improved fidelity under the detection of single-photon losses. Since reducing the number of elements is crucial for NISQ computing, our proposal provides an attractive way for quantum variational algorithms with bosonic qubits.

Our approach reduces the required elements, as it avoids the need for ancillary qubits when performing gate operations. While error detection utilizing the characteristics of the $02$ code necessitate ancillary qubits for parity measurement, a recent approach called virtual quantum error detection (VQED)~\cite{tsubouchi2023virtual} enables us to obtain the error detected expectation values without performing the parity measurement.

VQED is known to have the following advantages in achieving hardware-efficient computation at the NISQ stage. First, it can be executed with only one ancillary system and expectation value measurements on quantum bits and the ancillary system. Secondly, it can be used simultaneously with other error mitigation methods. Additionally, it is known to be robust against depolarizing errors on the ancillary system~\cite{tsubouchi2023virtual}. By combining this with our proposal, we may realize more hardware-efficient scheme, which is left for future work.

%%%%%%%%%
\begin{acknowledgments}
%%%%%%%%%
We are grateful to Tsuyoshi Yamamoto, Aiko Yamaguchi, Yohei Kawakami, Tomohiro Yamaji, Yuki Tanaka, Kohsuke Mizuno, Takaaki Aoki, Hiroshi Hayasaka, and Takashi Imoto for their insightful comments. This paper is partly based on the results obtained from a project, JPNP16007, commissioned by the New Energy and Industrial Technology Development Organization (NEDO), Japan. This work was supported by JST Moonshot  (Grant Number JPMJMS226C). Y. Matsuzaki is supported by JSPS KAKENHI (Grant Number 23H04390).
This work was also supported by CREST (JPMJCR23I5), JST.

We thank the developers of QuantumOptics.jl~\cite{kramer2018quantumoptics} which was used for our numerical simulations.

%%%%%%%%%
\end{acknowledgments}
%%%%%%%%%

\appendix
%%%%%%
\section{Second order perturbation}
\label{app:enederiv}
%%%%%%
We describe an effective Hamiltonian of our bosonic system. In the interaction picture, the Hamiltonian~\eqref{CZ:Ham} is described as
\begin{align}
    \hat{H}_{I}(t) = ge^{i\hat{H}_{0}t}(\hat{a}_{1}^{\dagger}\hat{a}_{2}+\hat{a}_{2}^{\dagger}\hat{a}_{1})e^{-i\hat{H}_{0}t},
\end{align}
where $\hat{H}_{0}$ is the first line of eq.~\eqref{CZ:Ham},
\begin{align}
   \hat{H}_{0}&= \sum_{i=1,2}\left(K_{i} \hat{a}_{i}^{\dagger 2}\hat{a}_{i}^{2} + \Delta_{i} \hat{a}_{i}^{\dagger} \hat{a}_{i}\right),\nonumber\\
   &= \sum_{i=1,2}\left(K_{i} \hat{n}_{i}^{2}+(\Delta_{i}-K_{i})\hat{n}_{i}\right).\label{Free_Ham}
\end{align}
Here, $\hat{n}_{i}=\hat{a}^{\dagger}_{i}\hat{a}_{i}$ is the number operator of the $i$-th KNRs. Using~\eqref{Free_Ham}, $\hat{H}_{I}(t)$ becomes
\begin{align}
    \hat{H}_{I}(t) &= ge^{it(2K_{1}(\hat{n}_{1}-1)-2K_{2}\hat{n}_{2}+\Delta_{1}-\Delta_{2})}\hat{a}_{1}^{\dagger}\hat{a}_{2} + h.c..\\
    &=g\hat{a}_{1}^{\dagger}\hat{a}_{2}e^{it(2K_{1}\hat{n}_{1}-2K_{2}(\hat{n}_{2}-1)+\Delta_{1}-\Delta_{2})}+h.c..
\end{align}

When we perform quantum computation for the states $|00\rangle$, $|02\rangle$, $|20\rangle$, and $|22\rangle$, the contribution of the first-order term in the perturbation arises as amplitudes for states outside the logical states. By maintaining an energy gap with states outside the logical space through Kerr coefficients $K_{i}$ and detuning parameters, \textit{i.e.}
\begin{align}
    K_{i},\Delta_{i},|\Delta_{i}-\Delta_{j}|\gg |g_{ij}|.
\end{align}
Using this condition, we can assume that amplitudes from the first-order perturbation do not substantially appear. Therefore, we will now proceed to calculate the second-order term. To obtain the effective Hamiltonian directly, we use the average Hamiltonian theory~\cite{Brinkmann_2016CMR}. 
\begin{widetext}
\begin{align}
    \hat{H}_{\rm eff}&=\frac{1}{2i(t_{f}-t_{i})}\int_{t_{i}}^{t_{f}} dt_{1}\int_{t_{i}}^{t_{1}} dt [\hat{H}_{I}(t_{1}),\hat{H}_{I}(t)],\nonumber\\
    &=-\frac{g^{2}}{2K_{1}\hat{n}_{1}-2K_{2}(\hat{n}_{2}-1)+\tilde{\Delta}}(\hat{n}_{1}+1)\hat{n}_{2}+\frac{g^{2}}{2K_{1}(\hat{n}_{1}-1)-2K_{2}\hat{n}_{2}+\tilde{\Delta}}\hat{n}_{1}(\hat{n}_{2}+1), \label{denominatornumber}
\end{align}
\end{widetext}
where $\tilde{\Delta}=\Delta_{1}-\Delta_{2}$. Here, we assume that the Hamiltonian $H_{0}$ has no degeneracy, and we drop high-frequency oscillating terms. In Eq.~\eqref{denominatornumber}, the number operator is seen in the denominator, and this can be defined as follows.
Let us define a function $f(\hat{n})$ that includes the number operator, and this satisfies the following
\begin{align}
    f(\hat{n})|n\rangle = f(n)|n\rangle,
\end{align}
for all $\hat{n}$ where $n$ is the eigenvalue of the number operator.
It is worth mentioning that, to avoid diversion in Eq.~\eqref{denominatornumber}, we should choose parameters to satisfy $(-2K_{1}n_{1}+2K_{2}(n_{2}-1)-\tilde{\Delta})\neq 0$ and $2K_{1}(n_{1}-1)-2K_{2}n_{2}+\tilde{\Delta}\neq 0$. Throughout the paper, we consider such paremeters.

These terms give us energy corrections
\begin{align}
    \Delta E_{02}(g) &= \langle 02|\hat{H}_{\rm eff}|02\rangle = \frac{2g^{2}}{2K_{2}-\tilde{\Delta}},\nonumber\\
    \Delta E_{20}(g) &= \langle 20|\hat{H}_{\rm eff}|20\rangle = \frac{2g^{2}}{2K_{1}+\tilde{\Delta}},\nonumber\\
    \Delta E_{22}(g) &=\langle 22|\hat{H}_{\rm eff}|22\rangle \nonumber\\
    &= -\frac{6g^{2}}{4K_{1}-2K_{2}+\tilde{\Delta}} -\frac{6g^{2}}{4K_{2}-2K_{1}-\tilde{\Delta}},\label{pert_hosei_ener}
\end{align}
This result is consistent with the effective Hamiltonian derived by a Schrieffer-Wolff transformation to the second order~\cite{Blais_2004PRA, Blais_2021RMP}.

Let us explain how to realize the parity measurements. We assume that an ancillary qubit is coupled with the KNR. The total Hamiltonian is given by Eq.~\eqref{Hamil_coup}. Then, we can derive the effective coupling Hamiltonian by using the average Hamiltonian theory~\cite{Brinkmann_2016CMR} as follows
\begin{align}
\hat{H}_{\rm I} &= \frac{g^{2}}{2K(\hat{n}-1)+\Delta-E}\hat{n}\ket{\downarrow}\bra{\downarrow}\nonumber\\
&\quad-\frac{g^{2}}{2K\hat{n}+\Delta-E}(\hat{n}+1)\ket{\uparrow}\bra{\uparrow}.\label{effective}
\end{align}
Since we consider a subspace spanned by $\ket{0}$, $\ket{1}$, and $\ket{2}$ for the KNR, we can rewrite this Hamiltonian as follows
\begin{align}
    \hat{H}_{\rm I, eff} = a_{0}+ a_{1}\hat{n}+a_{2}\hat{n}^{2}+ b_{0}\ket{\uparrow}\bra{\uparrow} + (b_{1}\hat{n}+b_{2}\hat{n}^{2})\ket{\uparrow}\bra{\uparrow},
\end{align}
where $a_{0}, a_{1}, a_{2}, b_{0}, b_{1}, b_{2}$ are determined such that the eigenvalues (and corresponding eigenvectors) of this Hamiltonian coincide with those of the Hamiltonian in Eq.~\eqref{effective} within the subspace.

Then, the term $(b_{1}\hat{n}+b_{2}\hat{n}^{2})\ket{\uparrow}\bra{\uparrow}$ works as an interaction between the ancillary qubit and KNR. To perform the parity measurement, it is necessary that a phase shift due to the interaction term for $|2\rangle $ is an even multiple of that for $|1\rangle $, \textit{i.e.}
\begin{align}
    2m(b_{1}+b_{2})=2b_{1}+2^{2}b_{2}.
\end{align}
This condition is satisfied if $b_{1}$ or $b_{2}$ is zero. Also, $b_{2}=0$ is equivalent to the condition $\Delta-E = -K$. Combining this condition with~\eqref{effective}, we obtain the following effective Hamiltonian,
\begin{align}
    \hat{H}_{\rm eff}&=\left(K+\frac{2g^{2}}{K}\right)\hat{n}^{2}+\left(\Delta-K-\frac{3g^{2}}{K}\right)\hat{n}\nonumber\\
    &\quad+\left(\frac{E}{2}+\frac{g^{2}}{2K}\right)\sigma_{Z}-\frac{2g^{2}}{K}\hat{n}\ket{\uparrow}\bra{\uparrow}.\label{effective_indmeas}
\end{align}

The last term $-\frac{2g^{2}}{K}\hat{n}\ket{\uparrow}\bra{\uparrow}$ provides the desired interaction described in Eq.~\eqref{det_Ham}. Although the other terms induce unwanted phase shift, we can correct this by using Z-rotation gates.

%%%%%%%%%%%%%%%%
\section{Logical single qubit rotations with detecting single-photon loss}
\label{appendix_B}
%%%%%%%%%%%%%%%%
Let us explain the effect of single-photon loss during single-qubit rotations. We assume that an initial state $\rho(0)$ is within an even parity photon number subspace. It is known that~\cite{MUeda_1989QOE, Chuang1997_PRA, Michael_2016PRX}, for the Lindblad equation,
\begin{align}
    \dot{\rho}=\gamma (\hat{a}\rho\hat{a}^{\dagger}-\frac{1}{2}\{\hat{a}^{\dagger}\hat{a},\rho\}),
\end{align}
the error channels for a given initial state $\rho (0)$ are analytically given by,
\begin{align}
    \rho(t)=\sum_{l=0} \hat{E}_{l}(t)\rho(0)\hat{E}^{\dagger}_{l}(t).
\end{align}
where 
\begin{align}
    \hat{E}_{l}(t)=\sqrt{\frac{(1-e^{-\gamma t})^{l}}{l!}}e^{-\frac{\gamma t}{2}\hat{n}}\hat{a}^{l}.
\end{align}

The state after projecting the state into even photon-number subspace is
\begin{align}
    \hat{P}\rho(t)\hat{P}=\sum_{l=0} \hat{E}_{2l}(t)\rho(0)\hat{E}^{\dagger}_{2l}(t).
\end{align}
In small $t$ region, this state is approximately equal to,
\begin{align}
    \hat{E}_{0}(t)\rho(0)\hat{E}^{\dagger}_{0}(t) = e^{-\frac{\gamma t}{2}\hat{n}}\rho(0)e^{-\frac{\gamma t}{2}\hat{n}}.\label{eq:errchan}
\end{align}
This means that, when we perform quantum error detection and no error is detected, the state is affected by the following error channel:
\begin{align}
    \hat{E}_{0}(t)=e^{-\frac{\gamma t}{2}\hat{n}}.\label{eq:0chan}
\end{align}
This dynamics can be described by the following non-Hermite Hamiltonian:
\begin{align}
    -i\frac{\gamma}{2}\hat{n}.\label{decay_term}
\end{align}
This means that our logical single-qubit rotation can be calculated by the original Hamiltonian with this additional non-Hetmitian term~\eqref{decay_term}. 

The effective Hamiltonian for the logical $Z$ rotation is described as
\begin{align}
    \hat{H}_{Z}=K\hat{n}^{2} + (\Delta-K-i\frac{\gamma}{2})\hat{n}.
\end{align}
By expanding this Hamiltonian with the basis of $|0\rangle$ and $|2\rangle$,we obtain the corresponding matrix representation as,
\begin{align}
    \begin{pmatrix}
        \langle 0|\hat{H}_{Z}|0\rangle & \langle 2|\hat{H}_{Z}|0\rangle \\
        \langle 0|\hat{H}_{Z}|2\rangle & \langle 2|\hat{H}_{Z}|2\rangle
    \end{pmatrix}
    =\begin{pmatrix}
        0 & 0 \\
        0 & 2K+2\Delta-i\gamma
    \end{pmatrix}.
\end{align}
The real part of the eigenvalue of the non-Hermite Hamiltonian represents the resonant energy while the imaginary part represents the decay rate~\cite{Dambrosio1998_CPV, Mori2021_PRA}.

On the other hand, when we consider the logical $X$ rotation, the effective Hamiltonian is represented by,
\begin{align}
    \hat{H}_{X} = K\hat{n}(\hat{n}-2) -i\frac{\gamma}{2}\hat{n} + p(\hat{a}^{2}+\hat{a}^{\dagger 2}),
\end{align}
and its matrix representation for the logical states is
\begin{align}
    \begin{pmatrix}
        \langle 0|\hat{H}_{X}|0\rangle & \langle 2|\hat{H}_{X}|0\rangle \\
        \langle 0|\hat{H}_{X}|2\rangle & \langle 2|\hat{H}_{X}|2\rangle
    \end{pmatrix}
    =\begin{pmatrix}
        0 & \sqrt{2}p \\
        \sqrt{2}p & -i\gamma
    \end{pmatrix}.
\end{align}
The eigenvalues of the Hamiltonian is
\begin{align}
\frac{-i\gamma\pm\sqrt{8p^{2}-\gamma^{2}}}{2}.
\end{align}
The angular frequency of the logical $X$ rotation is $\sqrt{8p^{2}-\gamma^{2}}$ and it means that the decay rate $\gamma$ leads to an effective delay of the logical $X$ rotation. It is known that when we perform the $X$ rotation, the dephasing channel $\hat{L}=\hat{a}^{\dagger}\hat{a}$ also brings similar change of the angular frequency~\cite{Kakuyanagi2016JPSJ}.

\bibliography{main}

\end{document}